# Effect of graphene and carbon-nitride nanofillers on the thermal transport properties of polymer nanocomposites: A combined molecular dynamics and finite element study

Leila Razzaghi[a], Maryam Khalkhali[a], Ali Rajabpour[b], Farhad Khoeini[a*]

[a] Department of Physics, University of Zanjan, Zanjan, 45195-313, Iran

[b] Advanced Simulation and Computing Lab. (ASCL), Mechanical Engineering Department, Imam Khomeini International University, Qazvin, 34148–96818, Iran

**ABSTRACT:** Low thermal conductivity of polymers, which is one of the considerable drawbacks of commonly used composite structures, has been the focus of many researchers aiming to achieve high-performance polymer-based nanocomposites through the inclusion of highly thermally conductive fillers inside the polymer matrices. Thus, in the present study, a multiscale scheme using non-equilibrium molecular dynamics (NEMD) and finite element (FE) method is developed to explore the impact of different nano-sized fillers (carbon-nitride and graphene) on the effective thermal conductivity of polyethylene-based nanocomposites. We show that the thermal conductivity of amorphous polyethylene at room temperature using the reactive bond order (REBO) interatomic potential is nearly $0.36 \pm 0.05$W/mK. Also, the atomistic results predict that, compared to the $C_3N$ and graphene nanosheets, the $C_2N$ nanofilm presents a much stronger interfacial thermal conductance (ITC) with polyethylene. Furthermore, the results indicate that the effective thermal conductivity values of $C_2N$-polyethylene, $C_3N$-polyethylene, and graphene-polyethylene nanocomposite, at constant volume fractions of 1%, are about 0.47, 0.56, and 0.74 W/mK, respectively. In other words, the results of our models reveal that the thermal conductivity of fillers is the dominant factor that defines the effective thermal conductivity of nanocomposites.

## 1. Introduction

In recent years, polymer nanocomposites have attracted significant interest among researchers because of their high potential applications in energy-related fields such as optoelectronics [1,2], thermoelectric [3,4], sensors [5,6], and batteries [7,8]. These polymer-based matrices not only benefit from the unique properties of the host material, such as high heat capacity, desirable chemical resistance, lightweight, stability, and nontoxicity, but also the addition of nano-sized fillers typically leads to improved thermal, mechanical, and electrical properties of the polymeric matrices [9–17].

Polymers are used in electronic devices such as Li-ion batteries because of their high capacity to absorb and release heat in phase changing procedure, thereby the temperature rise inside the battery pack is delayed, and the possibility of overheating decreases [18]. However, there are also drawbacks associated with commonly used polymers, including the low thermal conductivity (the thermal conductivity of pure amorphous polymers is typically in the range of 0.1-0.5 W/mK) [19], which is not desirable for the thermal management applications. One of the most appealing procedures to improve the thermal conduction features of polymers is combining polymers with the nano-sized materials with a much higher thermal conductivity [20–22]. Accordingly, extensive studies have been conducted on their physical properties, such as the thermal properties of polymer nanocomposites [23–26].

For instance, Vahedi et al. [27] investigated the effective thermal conductivity of CNT/paraffin nanocomposites by creating a multiscale scheme. They conducted molecular dynamics simulations to calculate the interfacial thermal conductance between CNT filler and surrounding paraffin. At the next stage, to explore the effective thermal conductivity of CNT/paraffin nanocomposites, they designed a representative volume element model of a macro-sized sample in the finite element method. Besides, they explored the effect of various geometric factors such as aspect ratio, volume fraction, and diameter on the effective thermal conductivity of nanocomposites. They observed that by increasing all the mentioned factors, the thermal conductivity of CNT/paraffin nanocomposites increases.

Mortazavi et al. [28] employed a multiscale method based on molecular dynamics simulations and the finite element approach to evaluate the effective thermal conductivity of graphene epoxy nanocomposites. They utilized molecular dynamics simulation to evaluate the thermal conduction of fillers and the matrix at the atomic scale. Also, they used a molecular dynamics approach to examine the thermal boundary conductance between graphene and epoxy. The results indicate that the thermal conductivity of graphene, acting as filler in the epoxy matrix, decreases by nearly 30%. Based on the MD results, they expanded the finite element method to explore the thermal conductivity of graphene epoxy nanocomposite. Also, they evaluated the impact of the formation of covalent bonds between fillers and polymer atoms on the effective thermal conductivity of graphene epoxy nanocomposites. Their results illustrated that the effective thermal conductivity of graphene epoxy nanocomposites declines by about 5% by the formation of covalent bonds between graphene and epoxy atoms.

As another example, Mortazavi et al. [29] employed an atomistic-continuum multiscale approach aiming to investigate the progress of the thermal management efficiency of the Li-ion batteries via utilizing the paraffin-based nanocomposites. They used Newman's pseudo-2D electrochemical model to simulate the electrochemical processes of a Li-ion battery. Besides, the effective thermal conductivity of paraffin-based nanocomposites, strengthened with graphene or h-BN nanofillers, was obtained by molecular dynamics/finite element multiscale method. Multiscale simulations illustrate that the thermal conductivity of h-BN/ paraffin nanocomposites, as well as its heat capacity, were higher than those of the graphene- paraffin nanocomposites with similar geometrical properties.

Among a few issues in the thermal transport behavior of polyethylene-based nanocomposites, various investigations concentrated on graphene-based fillers [30–32], and to date, limited research has been devoted evaluating the other types of nano-sized fillers. Owing to this fact, 2D carbon-nitride nanostructures, which are a new class of 2D materials, can be considered as efficient nano-sized fillers in polymeric matrices. It is worth mentioning that unlike graphene, 2D carbon-nitride nanostructures have a non-zero electronic energy bandgap, which makes them outstanding candidates for future applications of the next-generation electronic devices [33,34].

In the current study, a multiscale method is developed to investigate the impact of different nano-sized fillers (carbon-nitride and graphene) on the thermal conductivity of polyethylene-based nanocomposites. To this end, first, non-equilibrium molecular dynamics simulations are conducted to evaluate the thermal conductivity of amorphous polyethylene at the atomic scale. In the following step, the thermal relaxation method (pump-probe) is employed to calculate the interfacial thermal conductance between 2D nanostructures (carbon-nitride and graphene) and polyethylene. Finally, using the results obtained by the molecular dynamics simulations, finite element based three-dimensional models of the nanocomposite were constructed to explore the effective thermal conductivity at the microscale.

## 2. Methodology

### 2.1. Simulation details of polyethylene

Non-equilibrium molecular dynamics (NEMD) simulation was carried out using the Large Scale Atomic/Molecular Massively Parallel Simulator (LAMMPS) package [35] to compute the heat conductivity of amorphous polyethylene. To achieve this purpose, the initial simulation box was constructed in such a way that 96 polyethylene chains consist of 115392 atoms ($C_{400}H_{802}$ = 1202 atoms in a single chain) [36] were randomly and periodically placed inside a rectangular cubic box size of 220×73×36 Å. Reactive bond order (REBO) potential was used to describe atomic interactions between carbon (C) and hydrogen (H) atoms in polyethylene structure [37]. It is worth noting that Newton's equations of motion were integrated via the velocity Verlet algorithm [38] with a time step of 0.1 fs. In addition, the periodic boundary condition was employed in all directions.

In this simulation, firstly, the initial configuration experienced energy minimization to adjust atom coordinates. At the next stage, the whole system was relaxed at room temperature (300 K) for 1 ns under the NVE ensemble using the Langevin thermostat. Then, to impose the temperature gradient, and consequently heat flux as a response, the polyethylene box was divided into 22 slabs along the X-direction. To avoid rotations of the box during the simulation time, the outermost regions of the box were fixed. Adjacent to these fixed slabs, we placed hot and cold reservoirs set to 320 and 280

K, respectively, via the Nose-Hoover thermostat [39] under the NVT ensemble, while the remaining layers were imposed to constant energy (NVE) ensemble.

In the next step, when the system achieved non-equilibrium steady state heat transfer, the accumulative energy that added into and subtracted from polyethylene box was computed and plotted versus time. Accordingly, the heat current ($q_x$) was calculated based on the linear slopes of energy curves. Finally, the thermal conductivity of amorphous polyethylene was calculated from the well-known Fourier's formula in the X-direction as follows:

$$q_x = -\kappa A \frac{\partial T}{\partial X}, \quad (1)$$

where $q_x$ is the heat current, $\frac{\partial T}{\partial X}$ is the temperature gradient along the X-direction, and A is the cross-section area of the simulation box, which is perpendicular to the heat flux direction. It is notable that the system was simulated for the entire 1.3 ns after relaxation, and the first 0.9 ns were discarded as a pre-equilibration step. Molecular dynamics setup for evaluating the thermal conductivity of amorphous polyethylene has been shown in Fig. 1.

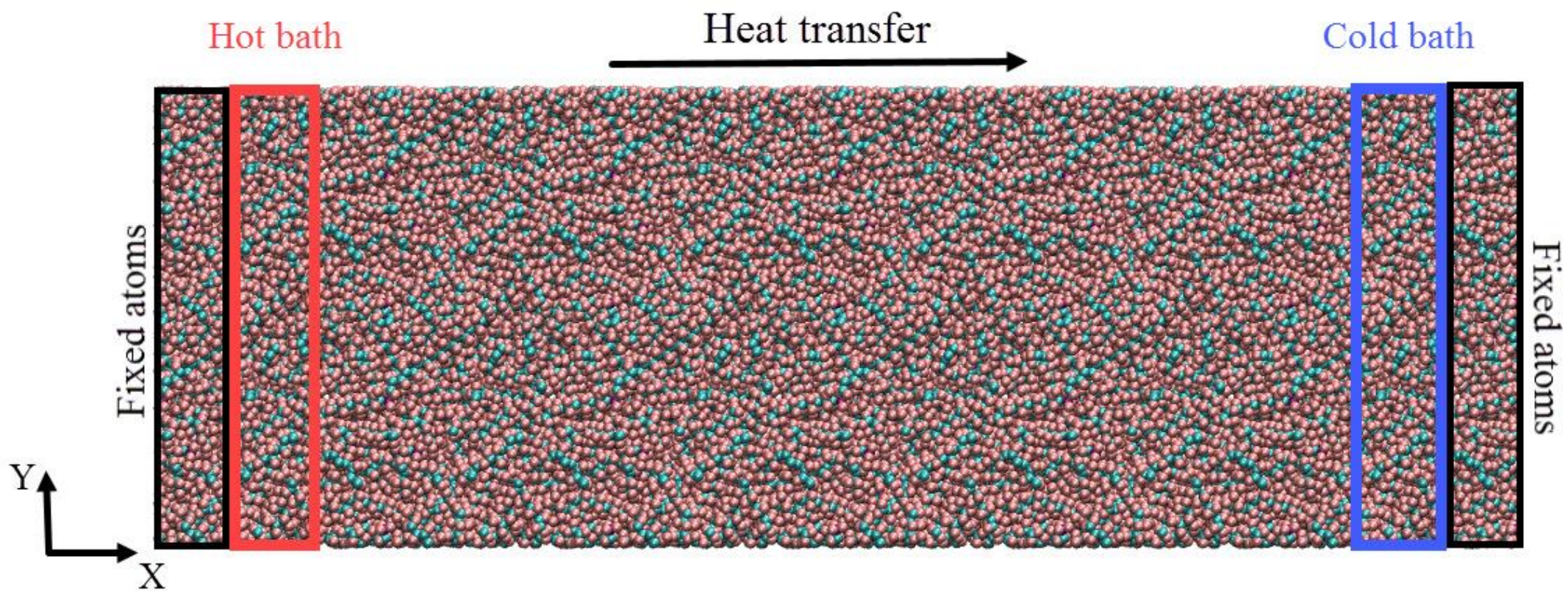

Fig. 1. Molecular dynamics setup for evaluating the thermal conductivity of amorphous polyethylene. The snapshot is captured from the polyethylene simulation box at the last time frame of the simulation. Carbon atoms are rendered in cyan and Hydrogen atoms in pink.

## 2.2. ITR between 2D nanomaterials (carbon-nitride and graphene) and polyethylene

Interfacial thermal resistance between 2D nanostructures (carbon-nitride and graphene) and polyethylene was evaluated using molecular dynamics simulation. To this end, the thermal relaxation method [40–42] was employed. The mentioned method is an MD approach, which is based on an experimental technique [43].

The Tersoff potential function is utilized to describe carbon (C) and nitrogen (N) interactions in 2D carbon-nitride nanostructures [44], as well as carbon-carbon interactions in graphene. As mentioned before, REBO potential was used to determine atomic interactions in polyethylene structure. Furthermore, the Lennard-Jones potential function was applied to describe nonbonding interactions between 2D nanostructures and polyethylene atoms. The Lennard-Jones coefficients are represented in Table 1.

Table 1. Lennard-Jones coefficients for van der Waals interactions between 2D nanostructures and polyethylene.

| | Ɛ (meV) | σ (Å) |
|---|---|---|
| $C_{polyethylene}$-$C_{carbon\text{-}nitride}$ | 2. 64 | 3.78 |
| $H_{polyethylene}$-$C_{carbon\text{-}nitride}$ | 1.60 | 3.27 |
| $C_{polyethylene}$-$N_{carbon\text{-}nitride}$ | 4.90 | 3.62 |
| $H_{polyethylene}$-$N_{carbon\text{-}nitride}$ | 2.60 | 3.34 |
| $C_{polyethylene}$-$C_{graphene}$ | 2.64 | 3.78 |
| $H_{polyethylene}$-$C_{graphene}$ | 1.60 | 3.27 |

## 2.3. Finite element modeling

Finally, the effective thermal conductivities of polyethylene nanocomposites at the microscale were evaluated using finite element modeling. To this purpose, we utilized the ABAQUS/Standard package (Version 6.14) and Python scripting.

Since the high computational costs of the finite element approach limits the modeling of composite structures, in the current study, the investigations of nanocomposites are limited to simulating the 3D cubic representative volume elements (RVE) with a restricted number of additives. Also, the geometry of 2D nanostructures (graphene, $C_3N$, and $C_2N$) was assumed to be disk-shaped. The diameter to thickness ratio was considered as the aspect ratio of the fillers.

In Fig. 2a, a specimen of created 3D cubic RVE model of polyethylene-based nanocomposite with 1% volume fraction of graphene, the $C_3N$, or the $C_2N$ platelets is represented. Due to the computational constraints, just 300 fillers with perfect disk-shaped geometry were randomly placed and dispersed inside the polymer matrix with no experience of the intersection with each other.

As shown in Fig. 2b, to evaluate the effective thermal conductivity of RVE along a particular direction, two thin films with the same segment size of the RVE box were put at both sides of the box to simulate the heat conductive surfaces. The thermal conductivity of the thin films was selected to be one million times that of the polymeric matrix. At this stage, a constant heat current was exerted on the RVE box. Consequently, a temperature difference as well as temperature gradient established along the heat flux direction, and the effective thermal conductivity of the sample was calculated using the one-dimensional form of the Fourier's law.

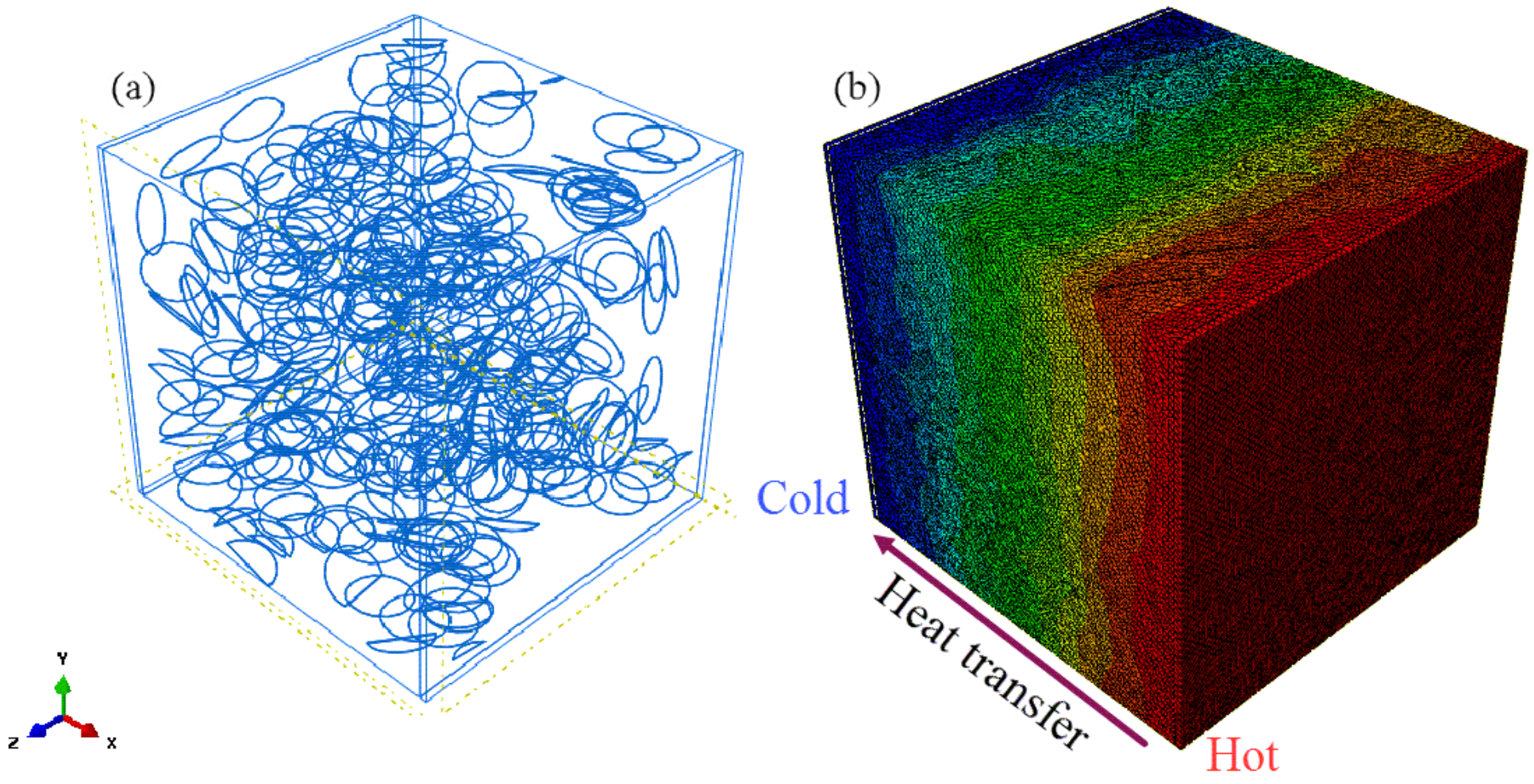

Fig. 2. (a) Finite element modeling of representative volume (RVE) of polyethylene nanocomposite representative volume element (RVE) with 1% concentration of 2D nanostructures (graphene, $C_3N$, and $C_2N$) nanofillers with an aspect ratio of 100. (b) 3D temperature profile for the finite element modeling of RVE of polyethylene nanocomposite.

## 3. Results and discussion

In this study, we developed a multiscale method consisting of atomistic molecular dynamics simulations and continuum modeling techniques to explore the effective thermal conductivity of amorphous polyethylene reinforced with graphene or 2D carbon-nitride nanosheet additives.

The steady-state temperature profiles of amorphous polyethylene specimen along X-direction is illustrated in Fig. 3a. According to this figure, by neglecting the nonlinearities near the two ends, which is caused by phonon scattering with the heat baths, one could observe a linear temperature gradient in the middle of the system. Considering the linear part of the temperature profile, the established slope is obtained as $|\frac{dT}{dX}| = 2.01\ (\frac{K}{nm})$ .

In Fig. 3b, we illustrate the accumulative added energy to the hot reservoir and the subtracted energy from the cold reservoir of the specimen. As depicted in Fig. 3b, the applied heat current $q_x = \frac{dE}{dt}$ is computed based on the slope of energy curves. Moreover, the amount of added energy to the hot layer is equivalent to the removed energy from the cold segment, which is the evidence of total energy conservation. Consequently, the thermal conductivity of amorphous polyethylene stood at nearly $0.36 \pm 0.05$W/mK employing the Fourier's law, which is in line with the results of previous studies [19,45].

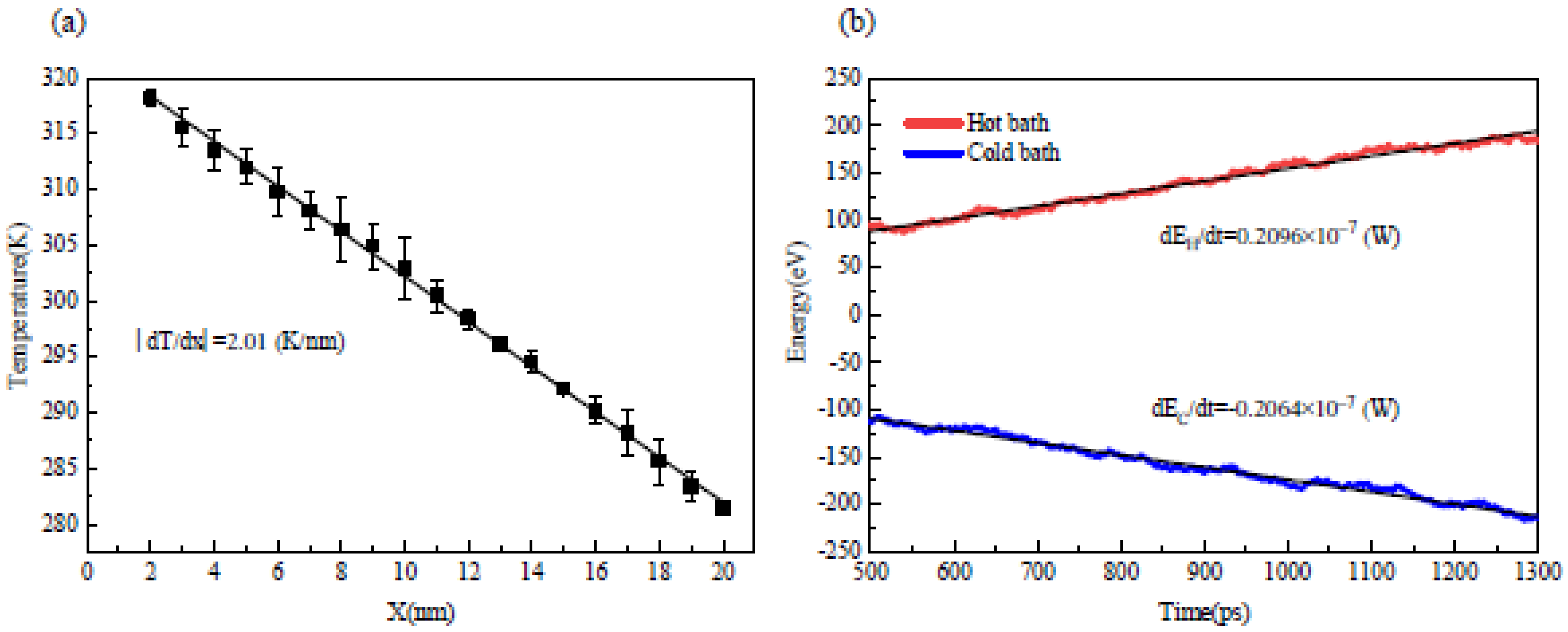

Fig. 3. (a) The steady-state temperature profiles of amorphous polyethylene specimen along X-direction due to the imposed temperature difference of ΔT = 40 K at T = 300 K. (b) Accumulative added energy to the hot region and subtracted energy from the cold area during the simulation time.

As discussed earlier, the thermal relaxation method is employed to acquire interfacial thermal resistance between 2D nanostructures (carbon-nitride and graphene) and amorphous polyethylene. The mentioned approach focuses on the dynamic thermal response of the sample and reduces the computational time, compared with non-equilibrium molecular dynamics. For this purpose, firstly, the amorphous polyethylene box was constructed, and a carbon-nitride, or the graphene sheet was assembled on the top of the polyethylene box. Then, the conjugate gradient method was utilized to minimize the energy of the system. To do so, the system was relaxed to atmospheric pressure at 300 K under the NPT ensemble for 300 ps. At the following stage, the NVE ensemble for 100 ps was exerted, and the equilibrium distance between 2D nanostructures and amorphous polyethylene was obtained nearly 3.3 Å. As illustrated in Fig. 4, while the system is under constant energy (NVE ensemble), a heat pulse of $\dot{q} = 7 \times 10^{-4}\ W$ was imposed on the 2D nanostructures for 50 fs. As regards exerting the heat pulse was quickly; an initial temperature difference was created between the 2D nanostructures and the polymer, and the temperature of the polymer remained constant at 300 K while that of nanosheets increased to nearly 560 K for graphene and the $C_3N$, as well as 700 K for the $C_2N$.

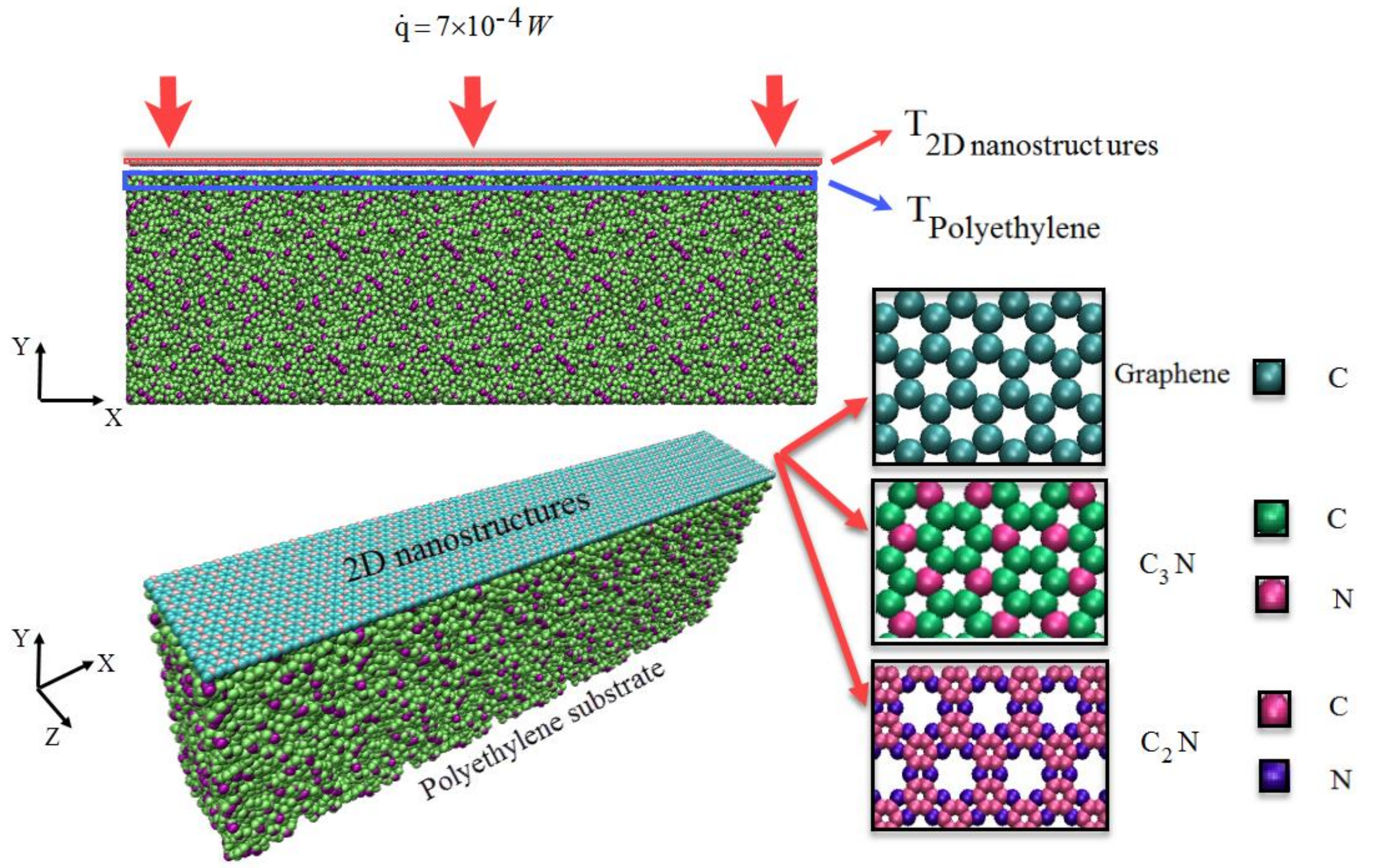

Fig. 4. Side and perspective view of the initial constructed atomistic models of 2D nanostructures (carbon-nitride or graphene) on amorphous polyethylene and the imposed heat pulse to the 2D nanostructures to calculate the interfacial thermal conductance.

Ultimately, the system was allowed to thermally relax at constant energy (NVE ensemble) by the heat transferred from 2D nanostructures to the polymeric substrate. The temperature of 2D nanostructures and the upper region of polyethylene, as well as the total energy of the 2D nanostructures, were calculated and recorded during the simulation process as a function of time. It should be noted that the recorded values for energy averaged over every 50 fs to suppress noises. The temperature and total energy variation of 2D nanostructures (graphene, $C_3N$, and $C_2N$) and amorphous polyethylene substrate that resulted from applying the heat pulse to reaching the equilibrium condition are depicted in Fig. 5a-c.

As depicted in Fig. 5a-c, the temperatures of 2D nanostructures (graphene, $C_3N$, and $C_2N$) and polyethylene were recorded during the relaxation procedure versus time. The temperature difference between the 2D nanostructures and the polymeric substrate decays exponentially.

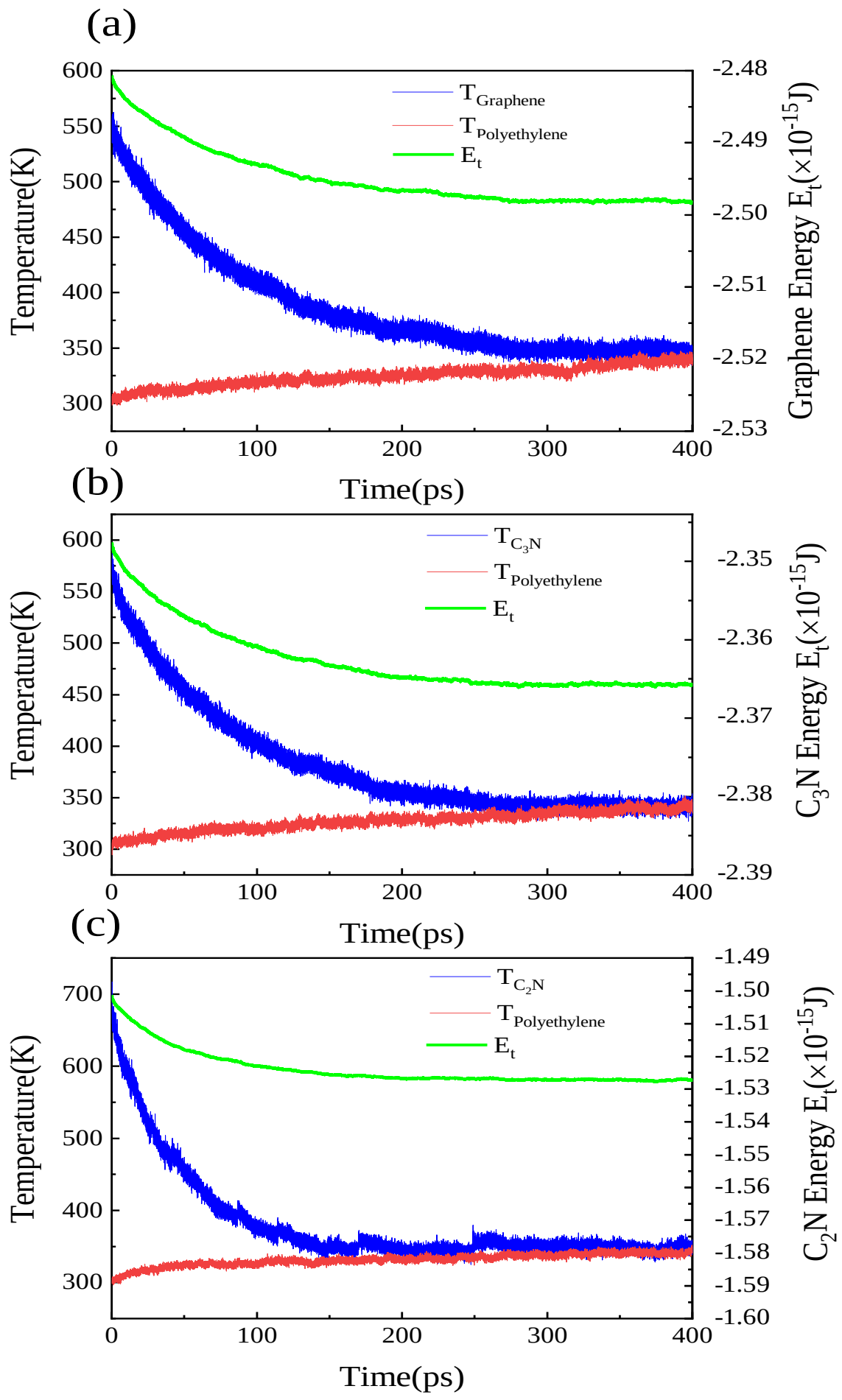

Fig. 5. Temperature and total energy evolutions for the evaluation of interfacial thermal resistance between (a) graphene and amorphous polyethylene, (b) the $C_3N$ and amorphous polyethylene, and (c) the $C_2N$ and amorphous polyethylene.

Utilizing the obtained MD results, the interfacial thermal resistance between 2D nanostructures (carbon-nitride and graphene) and amorphous polyethylene can be calculated by the following energy balance equation [46]

$$\frac{\partial E_t}{\partial t} = -\frac{A}{R}\left(T_{\text{2D nanostructure}} - T_{\text{polyethylene}}\right), \qquad (2)$$

where $E_t$ and $T_{\text{2D nanostructure}}$ refer to the total energy and temperature of the 2D nanostructures (graphene, $C_3N$, and $C_2N$), $T_{\text{polyethylene}}$ is the temperature of the polymeric substrate. R is the interfacial thermal resistance between 2D nanostructures and substrate, and A refers to the area through which the heat current was transferred.

By integrating Eq. (2) over time, we have

$$E_t = -\frac{A}{R}\int_0^t \left(T_{\text{2D nanostructure}} - T_{\text{polyethylene}}\right) dt + E_0. \qquad (3)$$

where $E_0$ is the initial total energy of the 2D nanostructures.
The total energy of 2D nanostructures with respect to the integral of the temperature difference between 2D nanostructures and polymeric substrate in each time step was recorded. The slopes of the linear fitting to the diagram, considering the cross-section area through which heat current transfers, yields the interfacial thermal resistance.
The results of the interfacial thermal conductance between different 2D nanostructures (graphene, $C_3N$, and $C_2N$) and amorphous polyethylene substrate at T = 300 K are depicted in Fig. 6. The ITC between 2D nanostructures (graphene, $C_3N$, and $C_2N$) and polyethylene was found to be 27, 31, and 53 MW/m$^2$K, respectively. The error bars were obtained through five simulations with different initial conditions, and also the standard deviation is used. Considering the thermal conductivity of 2D nanostructures, it is observed that the 2D structure with lower thermal conductivity had higher interfacial thermal conductance with polyethylene. It is worth mentioning that in this study, the thermal conductivity of graphene is considered to be 3000 W/mK, and the thermal conductivity of the $C_3N$ and the $C_2N$ is set to 62% and 4% that of graphene [46], respectively. To the best of our knowledge, there are no theoretical and experimental measurements for the interfacial thermal conductance between the $C_3N$ and the $C_2N$ with polyethylene. It can be said that from an experimental standpoint, the interfacial thermal conductance between different structures was often reported to be less than 50 MW/m$^2$K [47].

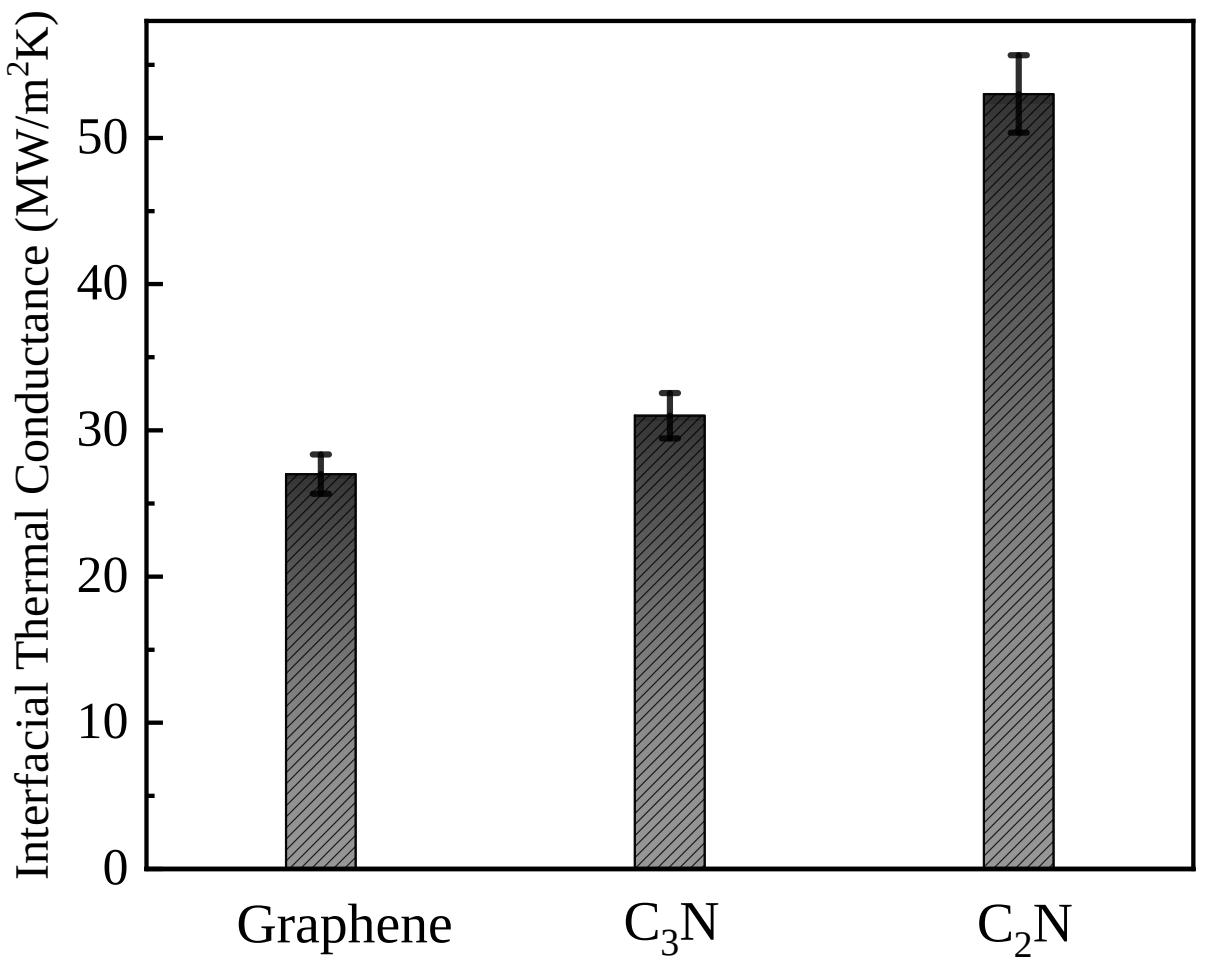

Fig. 6. The interfacial thermal conductance between different 2D nanostructures (graphene, $C_3N$, and $C_2N$) and amorphous polyethylene substrate at T = 300 K. The error bars are 3, 5, and 7 MW/m$^2$K for graphene, the $C_3N$, and the $C_2N$, respectively.

To better understand the differences in the interfacial thermal conductance value for various 2D nanostructures (graphene, $C_3N$, and $C_2N$) with polyethylene substrate, the phonon power spectral density of the structures is illustrated in Fig. 7.
The phonon density of states was acquired by calculating the Fourier transform of the autocorrelation function of the velocity of atoms belonging to the 2D nanostructures (graphene, $C_3N$, and $C_2N$) and also polyethylene substrate as follows [48–50],

$$P(\omega) = \sum_j \frac{m_j}{k_B T} \int_0^\infty e^{-i\omega t} < \boldsymbol{v}_j(t).\boldsymbol{v}_j(0) > dt, \quad (4)$$

where $m_j$ , $\boldsymbol{v}_j$, and $\omega$ are the mass and the velocity of atom j, and the angular frequency, respectively.

It is found that the phonon densities of states of graphene is in acceptable agreement with those of previous studies. Also, it is observed that there existed significant dissimilarity between the spectrum of each sheet with their polymeric substrates. As it is clear, if there is lower phonon coupling between two structures, and correspondingly, a remarkable difference between the phonon spectra, there will be a lower interfacial thermal conductance [46]. It is interesting to quantify the overlap between the two DOS curves. A lower overlap means that there is more phonon scattering at the interface of them and vice versa. The overlap is defined with the following equation [51,52],

$$H = \frac{\int_0^\infty D_1(\omega) D_2(\omega) d\omega}{\int_0^\infty D_1(\omega) d\omega \int_0^\infty D_2(\omega) d\omega} \quad (5)$$

where D1 and D2 are the DOS of two structures. The H values are represented in Table. 2 for three nano-fillers.

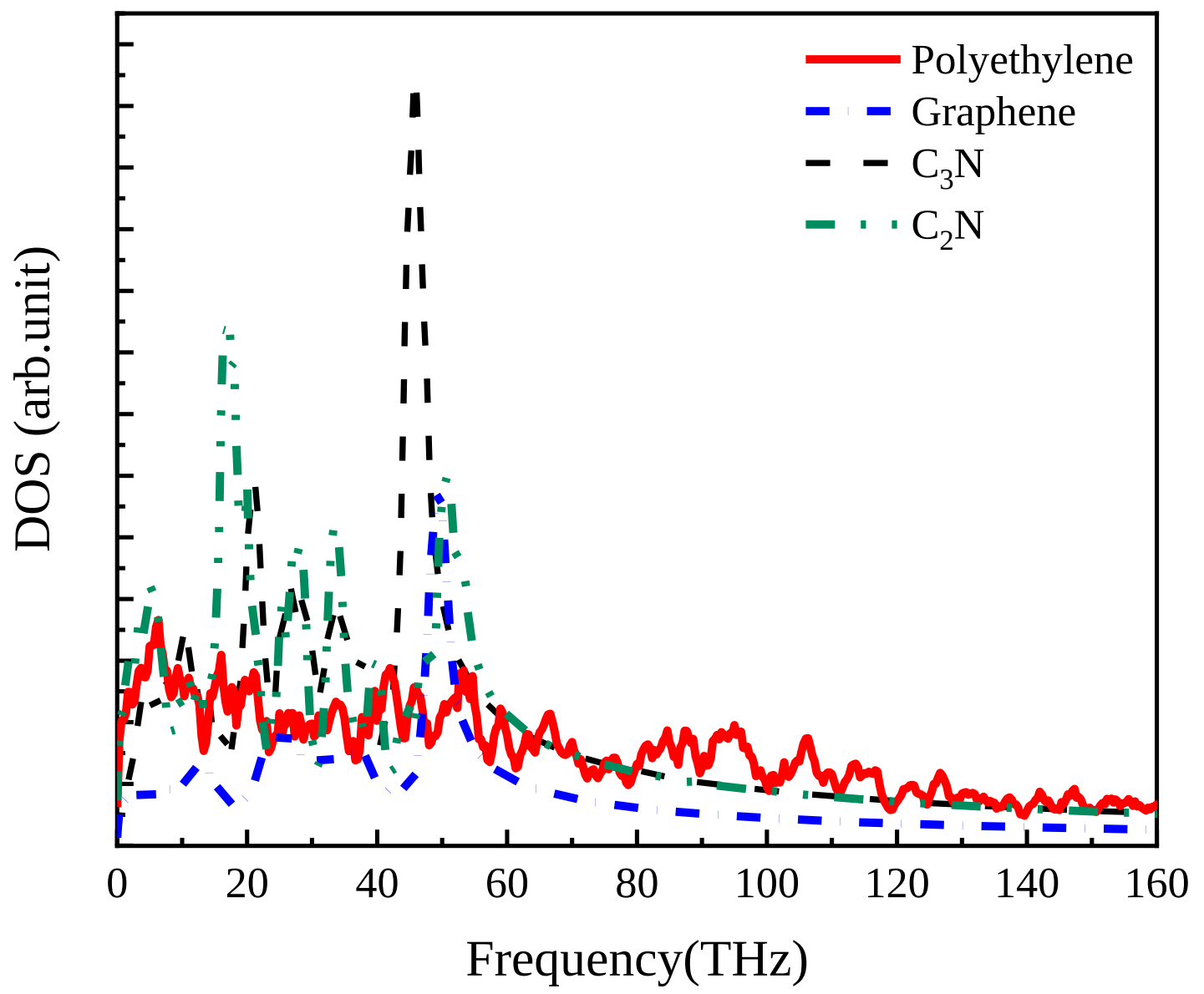

Fig. 7. Phonon densities of states of 2D nanostructures (graphene, $C_3N$, and $C_2N$) and the polyethylene substrate.

Table 2. The calculated overlap H for three nano-fillers with the matrix.

| Composite | Polymer-graphene | Polymer-$C_3N$ | Polymer-$C_2N$ |
|---|---|---|---|
| H | 0.0089 | 0.0092 | 0.0096 |

These values indicate that the amount of overlap is high for those with high thermal conductivity and vice versa. These values are in good agreement with thermal conductance of the systems.
In this section, the multiscale modeling results of the effective thermal conductivity of polyethylene nanocomposites are discussed. In the finite element approach of the nanocomposite, the thermal conductivity of $0.36 \pm 0.05$W/mK, which is obtained by our NEMD simulations, was assigned for polyethylene. Also, the thermal boundary conductance values acquired by the thermal relaxation method (Fig. 6) are utilized to establish the contact conductance between fillers and polymer.
The normalized effective thermal conductivity of polyethylene nanocomposites consists of different types of additives (graphene, $C_3N$, and $C_2N$) versus the thermal conductivity of fillers that is illustrated in Fig. 8. Besides, we studied the effect of the aspect ratio of fillers at constant volume fractions of 1% on the thermal conductivity of nanocomposites. The results indicate that the effective thermal conductivity values of $C_2N$-polyethylene, $C_3N$-polyethylene, and graphene-polyethylene nanocomposite at constant volume fractions of 1% are about 0.47, 0.56, and 0.74 W/mK, respectively. Therefore, according to Fig. 8, after normalization by the thermal conductivity of amorphous polyethylene, the obtained results are just about 1.30, 1.55, and 2.05 for $C_2N$-polyethylene, $C_3N$-polyethylene, and graphene-polyethylene nanocomposite, respectively.
As we have expected, higher thermal conductivity of an additive leads to a higher thermal transport through the nanocomposites and, consequently, a higher effective thermal conductivity of the polymer-based nanocomposites. Therefore, polyethylene nanocomposite with graphene fillers has the highest effective thermal conductivity in comparison with carbon-nitride fillers. Also, by decreasing the aspect ratio of fillers at a constant volume fraction, we observed a decrement in the effective thermal conductivity. By reducing the aspect ratio of platelets from 100 to 20, the effective thermal conductivity of nanocomposites decreases about 20%, 32%, and 34% for $C_2N$-polyethylene, $C_3N$-polyethylene, and graphene-polyethylene nanocomposite, respectively.
This reduction is because of the decrease in the contacting areas of 2D nanostructures (graphene, $C_3N$, and $C_2N$) and polyethylene, since the heat is transported between the interface zone of 2D nanostructures and polymeric matrix through their contacting areas [27,29].

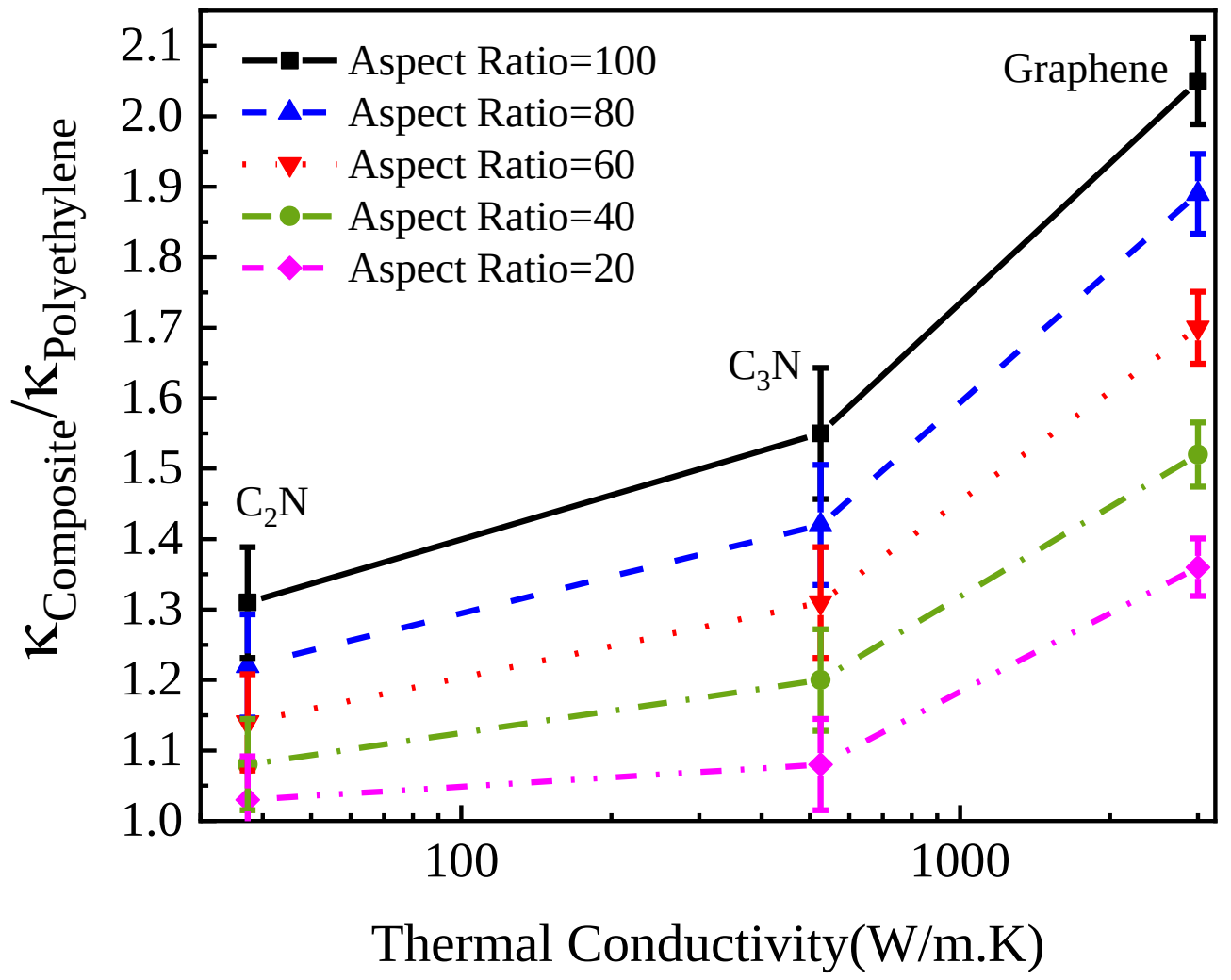

Fig. 8. The effective thermal conductivity of polyethylene nanocomposites consists of different types of additives at the distinct aspect ratio of fillers versus the thermal conductivity of fillers at a constant volume fraction of 1% (normalized by the thermal conductivity of amorphous polyethylene).

Another factor that may substantially affect the effective thermal conductivity of polyethylene nanocomposites, is the interfacial thermal conductance between 2D nanostructures (carbon-nitride and graphene) and polyethylene matrix. In Fig. 9, the effect of the interfacial thermal resistance between different 2D nanostructures and polymer host on the effective thermal conductivity of polyethylene-based nanocomposites at the distinct aspect ratio of fillers is explored. As it is shown, although the interfacial thermal conductance value between $C_2N$ and polyethylene is more than that of the $C_3N$ and graphene, the $C_2N$-polyethylene nanocomposite has the least effective thermal conductivity. This suggests that, among key factors that assign the effective thermal conductivity of polyethylene nanocomposites, the thermal conductivity of fillers has a more significant effect in comparison with the interfacial thermal conductance between additives and matrix. Also, as discussed before, by increasing the aspect ratio of additives at a constant volume fraction, an increment in the effective thermal conductivity of nanocomposite is found.

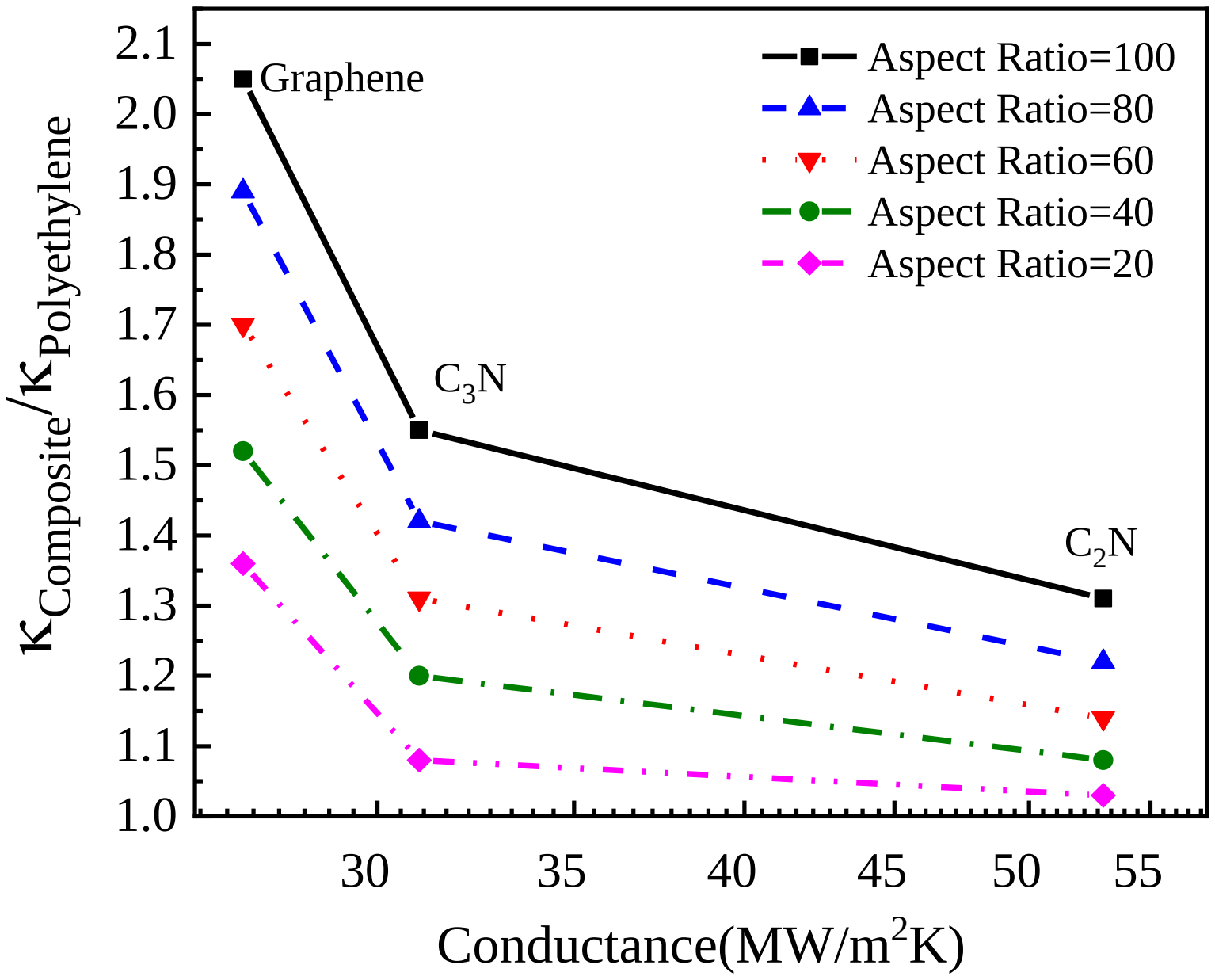

Fig. 9. The impact of interfacial thermal conductance between different 2D nanostructures and polymer host on the effective thermal conductivity of polyethylene-based nanocomposites at the distinct aspect ratio of fillers (normalized by the thermal conductivity of amorphous polyethylene).

In the last step, we explored the impact of nanofillers (graphene, $C_3N$, and $C_2N$) on the effective thermal conductivity of polyethylene nanocomposites. The normalized thermal conductivity of polyethylene-based nanocomposites at the different volume fractions of fillers versus the thermal conductivity of fillers is depicted in Fig. 9. It is notable that the effect of fillers volume fraction on the effective thermal conductivity of nanocomposites is evaluated at a constant aspect ratio of 100.

The key point that arose from Fig.10 is that by increasing the volume fraction of nanofillers from 1% to 3%, the normalized thermal conductivity of polymer nanocomposites increases by approximately 36%, 29%, and 23% for $C_2N$-polyethylene, $C_3N$-polyethylene, and graphene-polyethylene nanocomposite, respectively. As expected, the greater contribution of highly thermally conductive fillers within the polymer matrices leads to higher performance polymer-based nanocomposites [27].

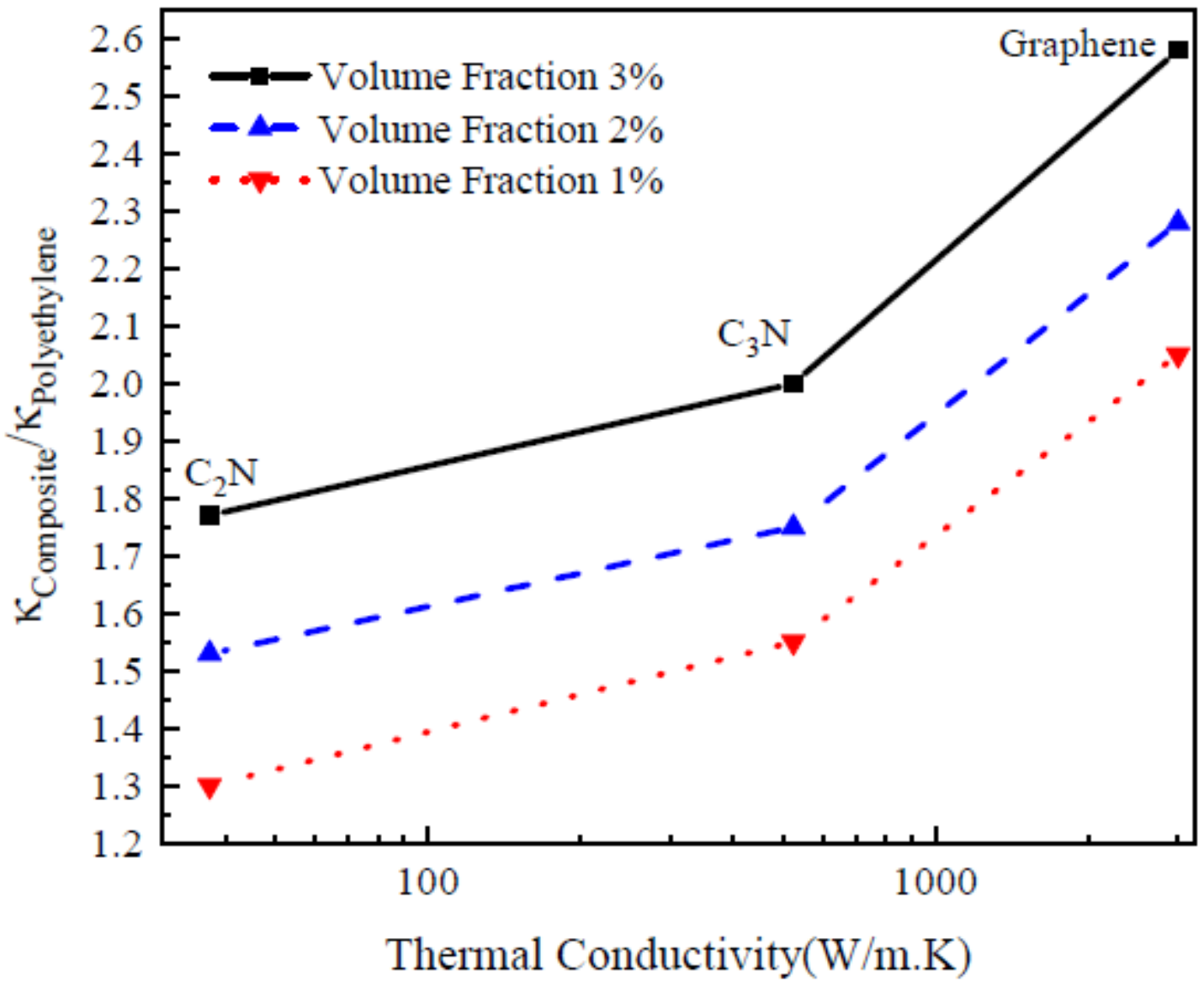

Fig. 10. The effective thermal conductivity of polyethylene nanocomposites consists of different types of additives at the distinct volume fraction of fillers versus the thermal conductivity of fillers at a constant aspect ratio of 100. (Normalized by the thermal conductivity of amorphous polyethylene).

## 4. Conclusions

In this study, we investigated the impact of different nano-sized fillers (carbon-nitride and graphene) on the effective thermal conductivity of polyethylene-based nanocomposites by performing multiscale modeling techniques.

Extensive NEMD simulations were conducted to explore the thermal conductivity of amorphous polyethylene at the atomic scale. Our results indicate that the thermal conductivity of amorphous polyethylene at room temperature was estimated to be almost $0.36 \pm 0.05$W/mK through utilizing reactive bond order (REBO) interatomic potential, which is in line with previous results in the literature.

In the next step, employing the thermal relaxation method, which is an MD technique inspired by an experimental approach, the interfacial thermal conductance between 2D nanostructures (graphene, $C_3N$, and $C_2N$) and polyethylene was evaluated. Our atomistic results show that the ITC between 2D nanostructures (graphene, $C_3N$, and $C_2N$) and polyethylene are nearly 27, 31, and 53 MW/$m^2$K, respectively. Therefore, considering the thermal conductivity of 2D nanostructures, obviously, the 2D structure with lower thermal conductivity has a higher interfacial thermal conductance with polyethylene, and the $C_2N$ nanofilm presents much stronger ITC with polyethylene, compared to the $C_3N$ and graphene nanosheets. Also, the underlying mechanism is demonstrated by calculating the phonon power spectral density.

Moreover, based on the results obtained by the molecular dynamics simulations, finite element based representative volume elements were constructed to evaluate the effective thermal conductivity of the nanocomposite. It is observed that the effective thermal conductivity values of $C_2N$-polyethylene, $C_3N$-polyethylene, and graphene-polyethylene nanocomposites at constant volume fractions of 1% were about 0.47, 0.56, and 0.74 W/mK, respectively. Polyethylene nanocomposite with graphene fillers had the most effective thermal conductivity in comparison with carbon-nitride fillers. In other words, the modeling results reveal that the dominant factor which defines the effective thermal conductivity of nanocomposites is the thermal conductivity of fillers, and ITC between 2D nanostructures (graphene, $C_3N$, and $C_2N$) and polyethylene plays a less important role for the heat transfer in the nanocomposites.

We also studied the impact of the aspect ratio of fillers at constant volume fractions on the thermal conductivity of nanocomposites. We found that by decreasing the aspect ratio of fillers at a constant volume fraction of 1%, the effective thermal conductivity decreased.

Finally, we explored the impact of volume fraction of nanofillers with constant aspect ratio on the thermal conductivity of Polyethylene nanocomposite. We observed that by increasing the volume fraction of additives from 1% to 3%, the normalized thermal conductivity of nanocomposites increases nearly 36%, 29%, and 23% for $C_2N$-polyethylene, $C_3N$-polyethylene, and graphene-polyethylene nanocomposite, respectively.

## Appendix A

In the context of the paper, we tried to describe the NEMD simulation and thermal relaxation method as two atomistic insights to study the composites. To the best of our knowledge, this is the first step towards investigating the effect of carbon-nitride nanofillers ($C_2N$ and $C_3N$) on the thermal transport properties of polyethylene-based nanocomposites, which have been carried out by a multiscale approach, non-equilibrium molecular dynamics (NEMD) and finite element (FE) methods, so far.

In the present study, the interfacial thermal resistance between 2D nanostructures and polyethylene was evaluated using the thermal relaxation method (pump-probe) approach, which is a routine MD approach based on an experimental technique. To show the accuracy of the thermal relaxation method, we also calculated the ITR between the 2D nanostructures and amorphous polyethylene by NEMD simulation as follows: polyethylene chains placed inside a rectangular cubic box with a size of 111×226×37.7 Å, and the nanosheet ($C_2N$, $C_3N$, or graphene) was placed in the middle of the box. The whole system was relaxed at room temperature (300 K) for 1 ns under the NPT ensemble. To impose a temperature gradient, the box was divided into 22 slabs along the Y-direction. The hot and cold reservoirs were set to 320 and 280 K, respectively, under the NVT ensemble, while the remaining layers were imposed on the NVE ensemble. The system was simulated for a total of 3 ns, after relaxation. Molecular dynamics setup for evaluating the interfacial thermal resistance between 2D nanostructures and the polyethylene is shown in Fig. A1.

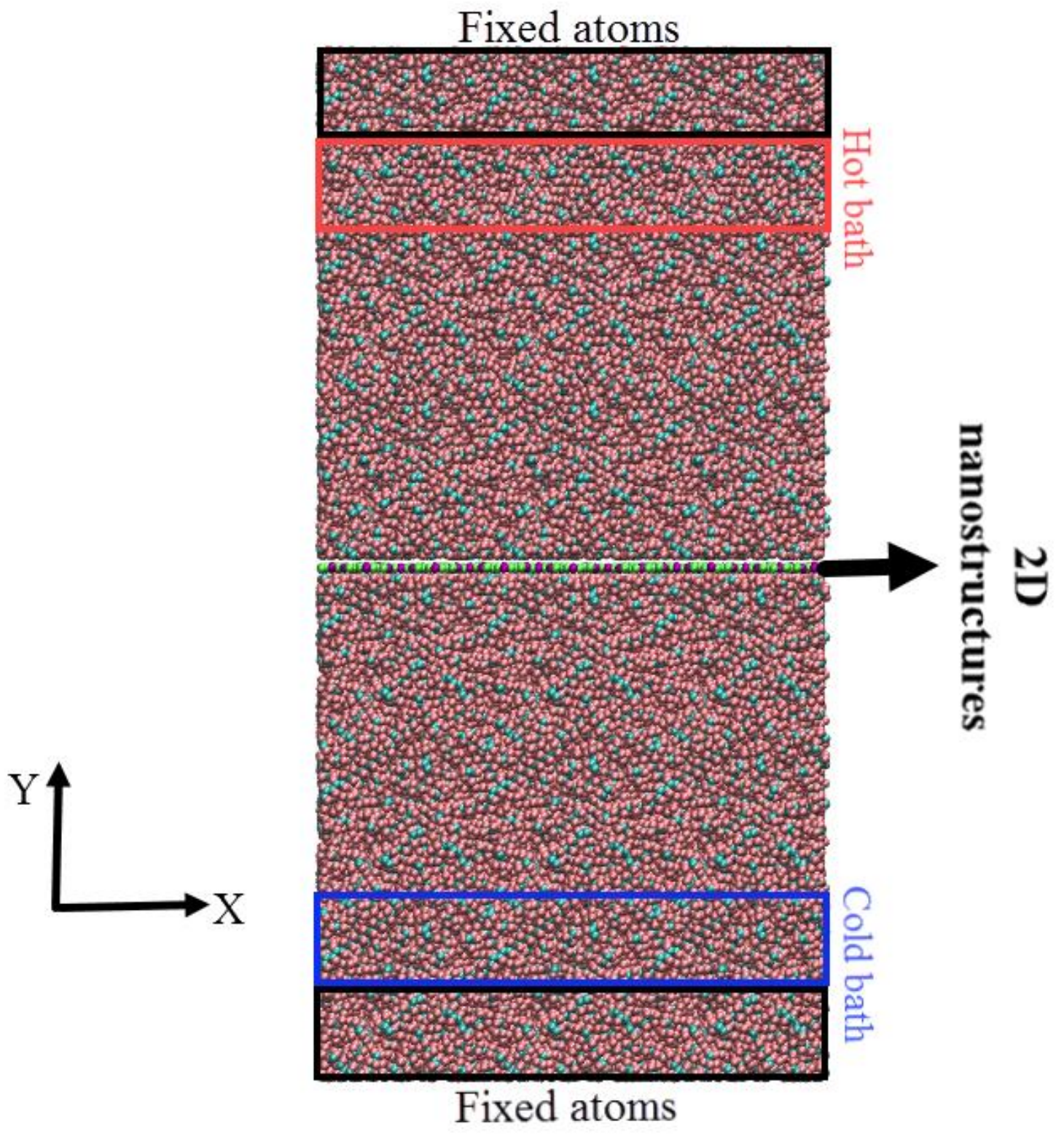

Fig. A1. Molecular dynamics setup for evaluating the interfacial thermal resistance between the 2D nanostructures and polyethylene.

In Fig. A2. (a), Accumulative added energy to the hot slab and subtracted energy from the cold slab as a function of the simulation time of the polyethylene box consisting of the graphene sheet is illustrated. The heat current (dE/dt) is computed as the slope of the linear fitted to energy profile. The steady-state 1D temperature profiles of the polyethylene box consisting of the graphene sheet along Y-direction is presented in Fig. A2. (b) A temperature jump of 9.4 K in the middle of the box due to the existence of the graphene sheet is observed.

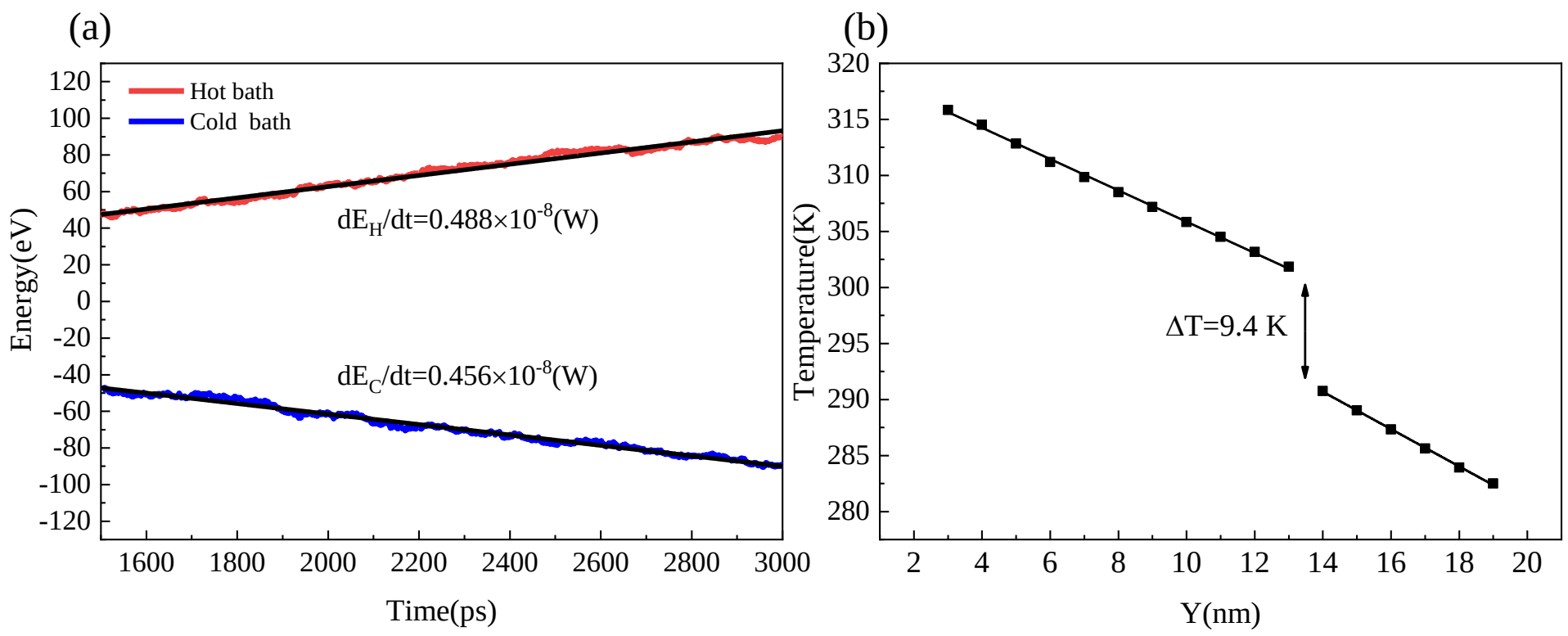

Fig. A2. (a) Accumulative added energy to the hot slab and subtracted energy from the cold slab as a function of the simulation time of the polyethylene box consisting of the graphene sheet. (b) The steady-state 1D temperature profiles of the polyethylene box consisting of the graphene sheet along Y- direction at T = 300 K and ΔT = 40 K.

The ITR and ITC of graphene/ polyethylene nanocomposite at room temperature are calculated as follows:

$$J = \frac{dE}{dt} \times \frac{1}{A} = \frac{0.0305 \times 1.6 \times 10^{-19}}{10^{-12}} \times \frac{1}{111 \times 37.7 \times 10^{-20}} = 1.16 \times 10^{8} \left(\frac{W}{m^2}\right) \qquad \text{(A.1)}$$

$$R = \frac{\left(\frac{\Delta T}{2}\right)}{J} = \frac{4.7}{1.16 \times 10^{8}} = 4.05 \times 10^{-8} \left(\frac{Km^2}{W}\right) \qquad \text{(A.2)}$$

$$G = \frac{1}{R} = 25 \left(\frac{MW}{Km^2}\right) \qquad \text{(A.3)}$$

In Fig. A3. (a), Accumulative added energy to the hot slab and subtracted energy from the cold slab as a function of the simulation time of the polyethylene box consisting of the $C_3N$ sheet is illustrated. The steady-state 1D temperature profiles of the polyethylene box consisting of the $C_3N$ sheet along Y-direction are presented in Fig. A3. (b) A temperature jump of 6.6 K in the middle of the box due to the existence of the $C_3N$ sheet is observed.

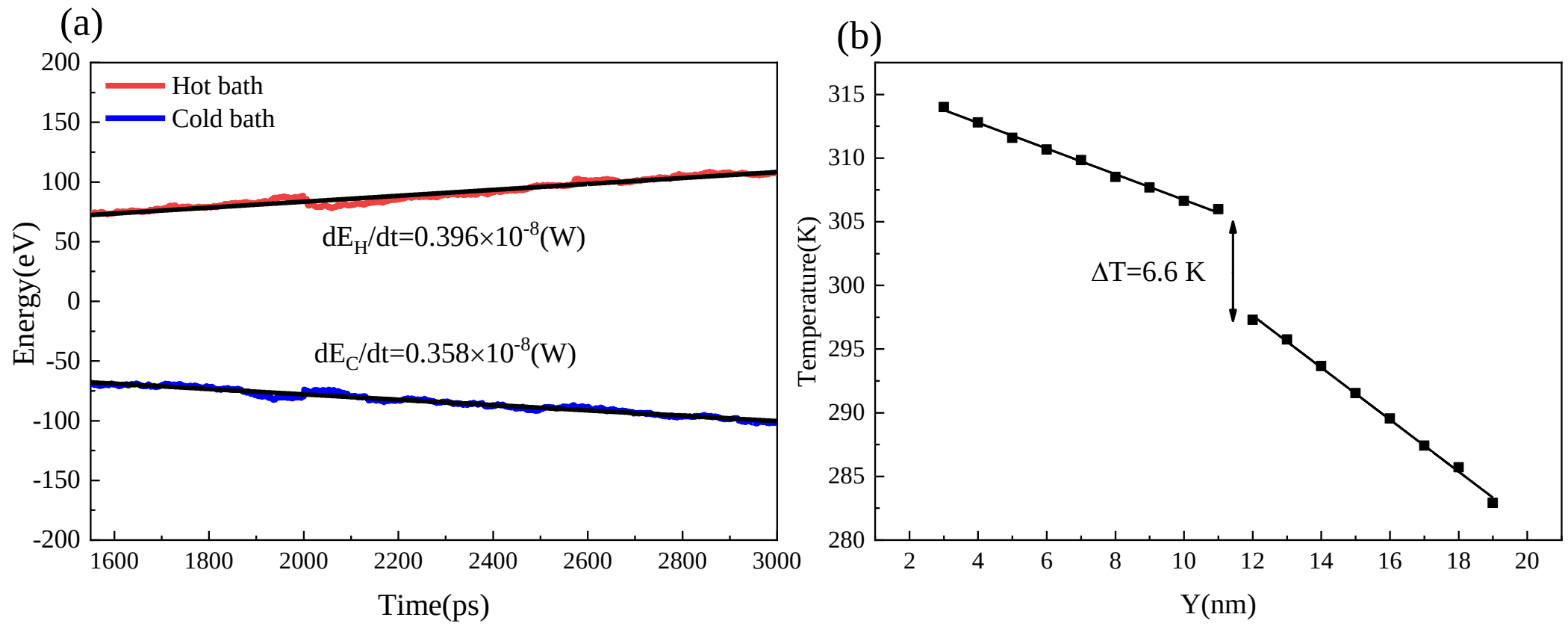

Fig. A3. (a) Accumulative added energy to the hot slab and subtracted energy from the cold slab as a function of the simulation time of the polyethylene box consisting of the $C_3N$ sheet. (b) The steady-state 1D temperature profiles of the polyethylene box consisting of the $C_3N$ sheet along Y-direction at T = 300 K and ΔT = 40 K.

The ITR and ITC of $C_3N$/polyethylene nanocomposite at room temperature are calculated as follows:

$$J = \frac{dE}{dt} \times \frac{1}{A} = \frac{0.0248 \times 1.6 \times 10^{-19}}{10^{-12}} \times \frac{1}{111 \times 37.7 \times 10^{-20}} = 0.94 \times 10^{8} (\frac{W}{m^2}) \qquad (A.4)$$

$$R = \frac{(\frac{\Delta T}{2})}{J} = \frac{3.3}{0.94 \times 10^{8}} = 3.51 \times 10^{-8} (\frac{Km^2}{W}) \qquad (A.5)$$

$$G = \frac{1}{R} = 28 (\frac{MW}{Km^2}) \qquad (A.6)$$

In Fig. A4. (a), Accumulative added energy to the hot slab and subtracted energy from the cold slab as a function of the simulation time of the polyethylene box consisting of $C_2N$ sheet is illustrated. The steady-state 1D temperature profiles of the polyethylene box consisting of $C_2N$ sheet along Y-direction are presented in Fig. A4. (b) A temperature jump of 3.4 K in the middle of the box due to the existence of $C_2N$ sheet is observed.

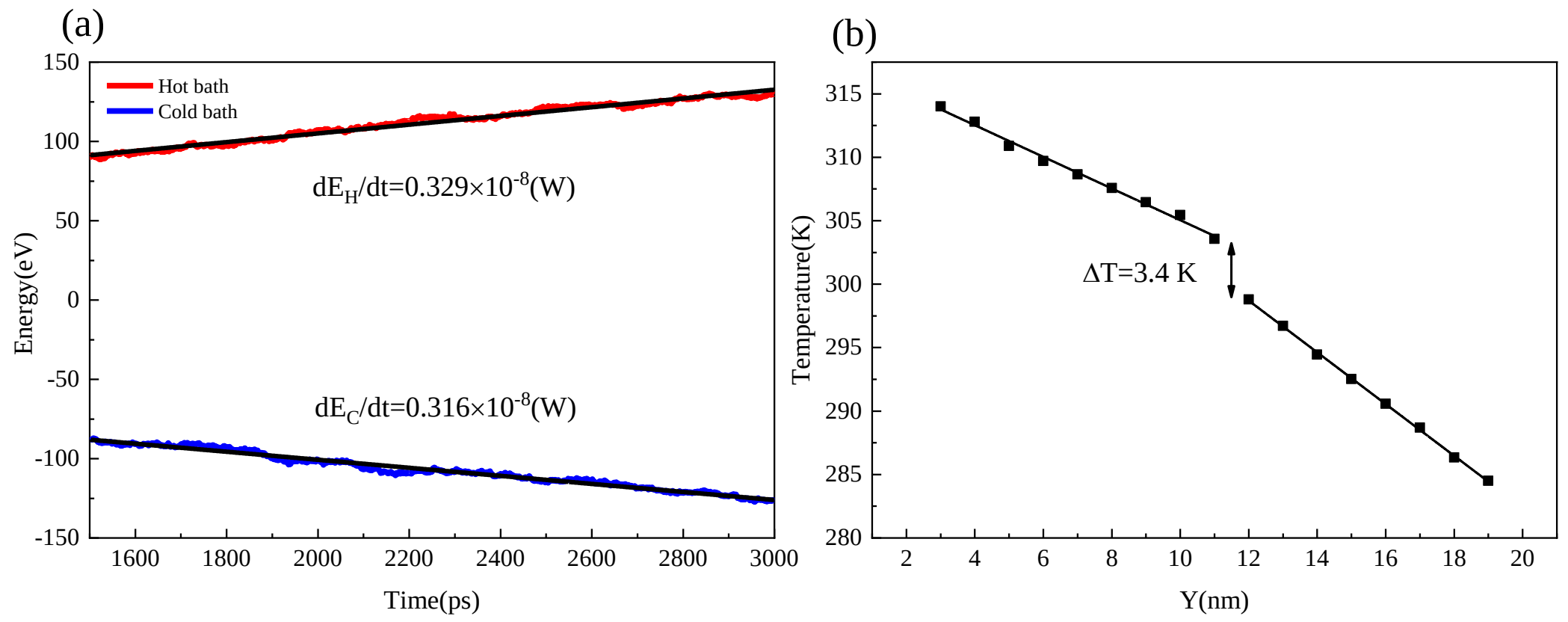

Fig. A4. (a) Accumulative added energy to the hot slab and subtracted energy from the cold slab as a function of the simulation time of the polyethylene box consisting of the $C_2N$ sheet. (b) The steady-state 1D temperature profiles of the polyethylene box consisting of the $C_2N$ sheet along Y-direction at T = 300 K and ΔT = 40 K.

The ITR and ITC of $C_2N$/polyethylene nanocomposite at room temperature are calculated as follows:

$$J = \frac{dE}{dt} \times \frac{1}{A} = \frac{0.0206\times1.6\times10^{-19}}{10^{-12}} \times \frac{1}{111\times37.7\times10^{-20}} = 0.78 \times 10^{8}(\frac{W}{m^2}) \qquad (A.7)$$

$$R = \frac{(\frac{\Delta T}{2})}{J} = \frac{1.7}{0.78\times10^{8}} = 2.18 \times 10^{-8}(\frac{Km^2}{W}) \qquad (A.8)$$

$$G = \frac{1}{R} = 46(\frac{MW}{Km^2}) \qquad (A.9)$$

Here, we compare the results of these methods. As shown in Table A1, the interfacial thermal conductance were obtained through both methods. The ITCs are close to each other in two methods. Therefore we can rely on both methods.

Table A1. The interfacial thermal conductance was obtained using NEMD and thermal relaxation methods.

| 2D nanostructures | ITC by thermal relaxation method ($MW/Km^2$) | ITC by Steady NEMD ($MW/Km^2$) |
|---|---|---|
| Graphene | 27 | 25 |
| $C_3N$ | 31 | 28 |
| $C_2N$ | 53 | 46 |

*Corresponding author: Farhad Khoeini (khoeini@znu.ac.ir)